\documentclass[12pt]{article}

\usepackage{amsmath}
\usepackage{amssymb}
\usepackage{amsfonts}
\usepackage{lscape}

\begin{document}

\fontsize{12}{6mm}\selectfont
\setlength{\baselineskip}{2em}

$~$\\[.35in]
\newcommand{\dss}{\displaystyle}
\newcommand{\raro}{\rightarrow}
\newcommand{\be}{\begin{equation}}

\def\sech{\mbox{\rm sech}}
\def\sn{\mbox{\rm sn}}
\def\dn{\mbox{\rm dn}}
\thispagestyle{empty}

\begin{center}
{\Large\bf Integrable Systems of}  \\    [2mm]
{\Large\bf Partial Differential Equations}  \\    [2mm]
{\Large\bf Determined by Structure Equations and Lax Pair}   \\   [2mm]
\end{center}

\vspace{1cm}
\begin{center}
{\bf Paul Bracken}                        \\
{\bf Department of Mathematics,} \\
{\bf University of Texas,} \\
{\bf Edinburg, TX  }  \\
{78541-2999}
\end{center}

\vspace{3cm}
\begin{abstract}
It is shown how a system of  evolution equations can be
developed both from the structure equations of a
submanifold embedded in three-space as well as
from a matrix $SO (6)$ Lax pair. The two systems obtained
this way correspond exactly when a constraint equation is
selected and imposed on the system of equations. 
This allows for the possibility of selecting the coefficients in the second
fundamental form in a general way.  
\end{abstract}

\vspace{2mm}
\vspace{2mm}

\newpage
{\bf 1. Introduction.}

There are some remarkable relationships between 
certain classes of partial differential equations and 
the geometry of surfaces, or submanifolds, immersed
in three-dimensional space which correspond to them {\bf [1]}.
Moreover, a great many partial differential
equations which are of interest to study and
investigate due to the role they play in various
areas of mathematics and physics are included in this
category {\bf [2-4]}. It must also be stated that these equations
typically result as the integrability condition of
a linear system or pair of linear equations usually referred 
to as a Lax pair. The cases of constant total or Gaussian
curvature as well as mean curvature have been studied
extensively. It might then be asked what can be said
with regard to the more general cases in which one or
both of the curvatures of the submanifold is or are
not constant.

Here we would like to ask what can be said about the
correspondence between partial differential equations
which can be obtained from a linear pair of matrix 
equations as well as from the structure equations
for a two-dimensional submanifold or surface which
is embedded in three-dimensional space such that the
coefficients of the second fundamental form
are left arbitrary at first.
This will turn out to produce a general
relationship between a system of partial differential
equations on the one hand and an associated surface
on the other. To state this another way, it is proposed
to see how previous results {\bf [5-6]} can be generalized 
to situations in which the curvatures of the submanifold
do not turn out to be identically constant. It will be 
found here that the moving frame approach will permit
the calculation of the basic fundamental forms which
are sufficient to determine the submanifold once the one-forms
in the structure equations are defined. It will be
seen that the system of equations which are produced
by the structure equations under a particular specification
of the basic one-forms can be exactly duplicated by
defining the form of a particular linear matrix Lax pair,
up to specifying a single constraint on some of the
functional quantities which appear.
These equations will be given for a specific choice
of one-forms, although
other choices may be possible. This work serves to
generalize the $SO (3)$ Lax pair which was produced
in {\bf [5]}. Generalizations of the $SO (2,1)$ Lax pairs
have also been done and will be reported later.

{\bf 2. Structure Equations and Differential Forms.}

Suppose ${\bf x} : M \rightarrow \mathbb R^3$ is a smooth
surface in $\mathbb R^3$. Choosing local coordinates
$t$ and $x$ in a coordinate neighborhood $U$ in $M$, the
surface can be expressed by the parametrized equations
$x^i = x^i (x,t)$, $ 1 \leq i \leq 3$. Choose a Darboux
frame $(x, e_1, e_2, e_3)$ on $M$ such that $e_1$ and
$e_2$ are tangent to $M$, $e_3$ is normal to $M$ and the
orientation of $(e_1, e_2, e_3 )$ is the same as a
chosen orientation of $\mathbb R^3$. Suppose the 
corresponding relative components for the frame field
are written $\omega_i$, $\omega_{ij}$, then {\bf [7-8]}
$$
d {\bf x} = \omega_1 e_1 + \omega_2 e_2,
\qquad
\omega_3 =0,
\eqno(2.1)
$$
$$
d e_j = \omega_{ji} e_i,
\qquad
\omega_{ij} + \omega_{ji} =0,
\eqno(2.2)
$$
where $\omega_i$, $\omega_{ij}$ are differential 
1-forms of the parameters $t$, $x$.
The structure equations are
$$
d \omega_1 = \omega_2 \wedge \omega_{21},
\qquad
d \omega_2 = \omega_1 \wedge \omega_{12},
\eqno(2.3)
$$
$$
\omega_1 \wedge \omega_{13} + \omega_2 \wedge \omega_{23}=0,
\eqno(2.4)
$$
$$
d \omega_{12} = \omega_{13} \wedge \omega_{32},
\qquad
d \omega_{13} = \omega_{12} \wedge \omega_{23},
\qquad
d \omega_{23} = \omega_{21} \wedge \omega_{13}.
\eqno(2.5)
$$
By Cartan's Lemma, based on (2.4), it follows that there exist
functions $h_{ij}$ such that
$$
\omega_{13} = h_{11} \omega_1 + h_{12} \omega_2,
\qquad
\omega_{23} = h_{21} \omega_{1} + h_{22} \omega_2,
\quad
h_{12} = h_{21} \equiv h.
\eqno(2.6)
$$
Now equations (2.5) are the Gauss equation and the
Codazzi equation of $M$. The first and second
fundamental forms can be obtained from the forms 
which appear in this system.
These essentially determine the surface up to rigid
motions. The first, second  
and third fundamental forms for $M$ are given by,
$$
I = d {\bf x} \cdot d {\bf x} = (\omega_1)^2 + ( \omega_2)^2,
\quad
II =- d {\bf x} \cdot d e_3 = \omega_1 \, \omega_{13}
+ \omega_2 \, \omega_{23},
\quad
III = d e_3 \cdot d e_3 = ( \omega_{13})^2 +
( \omega_{23})^2.
\eqno(2.7)
$$
The mean curvature and total curvature of $M$ are
determined by $h_{ij}$ and are both independent of
the choice of Darboux frame with
$$
H = \frac{1}{2} (  h_{11} + h_{22} ),
\qquad
K = h_{11} h_{22} - h_{12}^2.
\eqno(2.8)
$$
Now let us specify the forms, substitute them into 
the structure equations and simplify to see what
results without at first specifying the quantities
$h_{ij}$. When the coefficients of the forms are
subsequently given in terms of one or more unknown functions
$\varphi_j (x,t)$ as well, a system of partial differential
equations in terms of the $\varphi_j$ will be seen
to emerge,
$$
G_i ( \varphi_j, \varphi_{j,x}, \varphi_{j,t}, \cdots ) = 0.
\eqno(2.9)
$$
The key idea is that these partial differential equations will 
arise both from the structure equations
as well as from a matrix Lax pair. This means an important 
aspect of integrability is met automatically. Of course,
we keep to a pair of variables $x$, $t$ since the
Lax pair will depend on two variables. 
Keeping the level of complexity to a low level, 
a system of one-forms are defined $\omega_i$ which depend
on six functions $u_{ij}$ and $v_{ij}$ such that
the forms are given by
$$
\omega_1 = u_{12} \, dt + v_{12} \, dx,
\qquad
\omega_{2} = u_{13} \, dt + v_{13} \, dx,
\qquad
\omega_3 =0.
\eqno(2.10)
$$
It is the coefficients of the forms in (2.10),
$u_{ij}$ and $v_{ij}$, that depend on $\varphi_i$
which specify the differential equations.
The forms which specify the connection are 
$$
\omega_{12} = u_{23} \, dt + v_{23} \, dx,
\eqno(2.11)
$$
and $\omega_{13}$, $\omega_{23}$ which can be written
down using Cartan's Lemma given $\omega_1$ and $\omega_2$
from (2.10) are
$$
\omega_{13} = ( h_{11} u_{12} + h u_{13} ) \, dt
+ ( h_{11} v_{12} + h v_{13} ) \, dx,
\quad
\omega_{23} = ( h u_{12} + h_{22} u_{13} ) \, dt
+ ( h v_{12} + h_{22} v_{13} ) \, dx.
\eqno(2.12)
$$
Differentiating the forms and substituting into (2.3)
and (2.5), the results
can be summarized as follows
$$
\begin{array}{c}
u_{12,x} - v_{12,t} +u_{23} v_{13} - u_{13} v_{23} =0,  \\
  \\
u_{13,x} - v_{13,t} + u_{12} v_{23} - u_{23} v_{12} =0,  \\
  \\
u_{23,x} - v_{23,t} + (h_{11} h_{22} - h^2) ( u_{13} v_{12} - u_{12} v_{13}) =0, \\
  \\
( h_{11} u_{12} + h u_{13} )_x - (h_{11} v_{12} + h v_{13})_t
+ u_{23} (h v_{12} + h_{22} v_{13}) - v_{23} ( h u_{12} + h_{22} u_{13})   =0,  \\
  \\
( h u_{12} + h_{22} u_{13})_x - ( h v_{12} + h_{22} v_{13})_t
+ v_{23} (h_{11} u_{12} + h u_{13}) - u_{23} ( h_{11} v_{12}
+ h v_{13})=0.   \\
\end{array}
\eqno(2.13)
$$
This is the system of equations which results from the
structure equations.

{\bf 3. Equations Determined by a Linear System.}

It will now be seen how system of equations
(2.13) can be obtained from an $SO (6)$
matrix Lax pair. It is to be required that the two
linear systems
$$
\Phi_t = U \Phi,
\qquad
\Phi_x = V \Phi
\eqno(3.1)
$$
generate system (2.13) when the zero curvature
condition is enforced. The integrability condition which
follows from this set of linear equations (3.1) in terms
of $U$ and $V$ takes the form
$$
U_x - V_t + [ U, V] =0.
\eqno(3.2)
$$
It will be shown that there is at least one way to
get (2.13) by picking $U$, $V$ appropriately. 
Suppose we take $U = U_1 \oplus U_2$ and $V= V_1 \oplus V_2$
of the form
$$
U = \left(
\begin{array}{cc}
U_1  &  0   \\
0    &  U_2  \\
\end{array}  \right),
\qquad
V = \left(
\begin{array}{cc}
V_1  &  0   \\
0    &  V_2  \\
\end{array}  \right),
\eqno(3.3)
$$
so the compatibility condition (3.2) reduces to
$U_{i,x} - V_{i,t} + U_i V_i - V_i U_i =0$ for
$i=1,2$. To specify the submatrices $U_1$  and $V_1$, we take
the following $SO (3)$ matrices,
$$
U_1 = \left(
\begin{array}{ccc}
0  &  u_{12}  &  u_{13}  \\
-u_{12}  &  0  &  u_{23}  \\
- u_{13} & -u_{23}  &  0  \\
\end{array}   \right),
\quad
V_1 = \left(
\begin{array}{ccc}
0  &  v_{12}  &  v_{13}   \\
- v_{12}  &  0  &  v_{23}  \\
- v_{13}  &  - v_{23}  &  0  \\
\end{array}  \right).
\eqno(3.4)
$$
The submatrices $U_2$ and $V_2$ are given by
$$
U_2 = \left(
\begin{array}{ccc}
0  &  h_{11} u_{12} + h u_{13}  &  h u_{12} + h_{22} u_{13}  \\
- h_{11} u_{12} - h u_{13} &  0  &  u_{23}   \\
-h u_{12} - h_{22} u_{13}  & -u_{23}  &  0   \\
\end{array}  \right),
$$
$$
V_2 = \left(
\begin{array}{ccc}
0  &  h_{11} v_{12} + h v_{13}  &  h v_{12} + h_{22} v_{13}   \\
- h_{11} v_{12} - h v_{13}  &  0  &  v_{23}   \\
- h v_{12} - h_{22} v_{13}  & -v_{23}  &   0  \\
\end{array}  \right).
\eqno(3.5)
$$
Substituting (3.4), (3.5) into (3.3) and then using (3.3)
in (3.2), it is a straightforward calculation to show that (3.2)
is satisfied provided the following system holds,
$$
u_{12,x} - v_{12,t} + u_{23} v_{13} - u_{13} v_{23} =0,  \\
$$  
$$
u_{13,x} - v_{13,t} + u_{12} v_{23} - u_{23} v_{12} =0,  \\
$$
$$  
u_{23,x} - v_{23,t} + u_{13} v_{12} - u_{12} v_{13} =0,  \\
$$
$$
u_{23,x} - v_{23,t} + (h_{11} h_{22} - h^2) (u_{12} v_{12} - u_{12} v_{13}) =0,
\eqno(3.6)
$$
$$  
( h_{11} u_{12} + h u_{13})_x - ( h_{11} v_{12} + h v_{13} )_t
+ u_{23} ( h v_{12} + h_{22} v_{13}) - v_{23} ( h u_{12} + h_{22} u_{13})=0,  \\
$$  
$$
(h u_{12} + h_{22} u_{13})_x - ( h v_{12} + h_{22} v_{13})_t
+ v_{23} (h_{11} u_{12} + h u_{13}) - u_{23} (h_{11} v_{12} + h v_{13})=0,    \\
$$  
The matrices $U$ and $V$ therefore reproduce all five equations in
(2.13) which result from the structure equations except there is
a additional equation. The third and fourth equations in (3.6)  
appear in two different forms. In fact, these
two different forms can be exactly matched by introducing a
constraint, which can be chosen in two ways. These two
equations will match provided that
$$
 ( u_{12} v_{13} - u_{13} v_{12})( h^2 - h_{11} h_{22} +1) =0.
\eqno(3.7)
$$
Clearly, this can be satisfied in two different ways.
First of all, it can be satisfied by putting a constraint
on four of the functions $u_{ij}$ and $v_{ij}$, namely,
$u_{12} v_{13} - u_{13} v_{12} =0$.
Using this, no restrictions need be
placed on the coefficients of the second fundamental form.
Another way to satisfy (3.8) is to put a constraint on the
set of functions $h_{ij}$, namely,
$h_{11} h_{22} - h^2 =1$.
This result states that the total curvature
must be one, beyond that, the quantities 
appearing in it can be arbitrary. 

If the constraint on $u_{ij}$, $v_{ij}$ is substituted
in the third and sixth equations in (3.6), these
two equations reduce to the common expression
$ u_{23,x} - v_{23,t} =0$. Moreover,
$v_{13} = u_{13} v_{12}/u_{12}$ must be substituted
into the remaining equations in (3.6), and no external
restriction need be placed on the second fundamental form.
On the other hand, if we suppose the coefficient matrix of 
the second fundamental form has elements which satisfy $h_{11} h_{22} -h^2 =1$,
the third and sixth equations of (3.6) reduce to a
form with a common structure,
so these five equations exactly match (2.13). 
These five equations can be solved as a system.

Notice that these results imply that the pair of three by three
matrices which form the matrices (3.3) are not completely
independent of each other or completely decoupled at the
end.They are in a way coupled together by means of the
constraint which is imposed. Suppose $h_{11}=h_{22}=1$, 
then the final pair of equations in (2.13) go into
two of the first three equations. As a short example for this choice of
constraint, if we let $u_{13}=v_{12}=0$, $u_{12}= \cos (\varphi/2)$
and $v_{13} = \sin (\varphi/2)$ then (2.13)-(3.6) become
$$
(\cos (\frac{\varphi}{2}))_x + u_{23} \sin (\frac{\varphi}{2}) =0,
\quad
 - (\sin (\frac{\varphi}{2}))_t 
+ \cos (\frac{\varphi}{2}) v_{23} =0,
\quad
u_{23,x} - v_{23,t} - \cos (\frac{\varphi}{2}) \sin (\frac{\varphi}{2} ) =0.
$$
Solving for $u_{23}$ and $v_{23}$ from the first pair
and substituting into the third, we obtain that $\varphi$
satisfies the integrable equation 
$\varphi_{tt} - \varphi_{xx} =- \sin (\varphi)$.

%\vspace{2mm}
{\bf References.} \\
\noindent
$[1]$ C. Rogers and W. F. Schief, B\"acklund and Darboux
Transformations, Geometry and Modern Applications in
Soliton Theory, Cambridge Texts in Applied Mathematics,
Cambridge University Press, (2002).  \\
$[2]$ G. Chaohao, Soliton Theory and Its Applications,
Springer-Verlag, Berlin (1995).  \\
$[3]$ A. S. Fokas and I. M. Gelfand,
Surfaces on Lie Groups, on Lie Algebras, and Their Integrability,
Commun. Math. Phys.,
{\bf 177}, (1996), 203-220.  \\
$[4]$ A. S. Fokas, I. M. Gelfand, F. Finkel and Q. Liu,
A formula for constructing infinitely many surfaces
on Lie algebras and integrable equations,
Selecta Mathematica, New Series, {\bf 6}, (2000), 347-375.  \\
$[5]$ P. Bracken, Partial Differential Equations Which Admit
Integrable Systems, Int. J. of Pure and Applied Math., {\bf 43}, 
(2008), 408-421.  \\
$[6]$ P. Bracken, Integrable Systems Determined by Differential
Forms for Moving Frames on Immersed Submanifolds, Int. J.
Geometric Methods in Mod. Physics, {\bf 5}, (2000), 1041-1049.  \\
$[7]$ S. S. Chern, W. H. Chen and K. S. Lam, Lectures on
Differential Geometry, World Scientific, Singapore, (1999).  \\
$[8]$ H. Cartan, Differential Forms, Dover Pub., Mineola, NY, (2006).  \\
\end{document}